# Effective and Scalable Math Support: Experimental Evidence on the Impact of an AI- Math Tutor in Ghana

Owen Henkel[1], Hannah Horne-Robinson[2], Nessie Kozhakhmetova, Amanda Lee[3]

[1] University of Oxford
[2] Rising Academies
[3] J-PAL North America

**Abstract.** This study is a preliminary evaluation of the impact of receiving extra math instruction provided by Rori, an AI-powered math tutor accessible via WhatsApp, on the math performance of approximately 500 students in Ghana. Students assigned to both the control and treatment groups continued their normal classes with identical curricula and classroom hours. Students in the treatment group were given access to a phone for one hour a week – during their study hall period – and were allowed to use Rori to independently study math. All other aspects of the groups' in-school experience were the same. We find that the math growth scores were substantially higher for the treatment group and statistically significant ($p < 0.001$). The effect size of 0.36 is considered large in the context of educational interventions: approximately equivalent to an extra year of learning. Importantly, Rori works on basic mobile devices connected to low-bandwidth data networks, and the marginal cost of providing Rori is approximately $5 per student, making it a potentially scalable intervention in the context of LMICs' education systems. While the results should be interpreted judiciously, as they only report on year 1 of the intervention, they do suggest that chat-based tutoring solutions leveraging AI could offer a cost-effective and operationally efficient approach to enhancing learning outcomes for millions of students globally.

## 1 Introduction

Fewer than 15% of students in Africa achieve minimum proficiency in math by the end of middle school [1]. Most students in the region are taught content that surpasses their ability level, have limited opportunities to practice new skills, and often do not receive the necessary feedback due to large class sizes and under-resourced teachers [1]. Research has long suggested that high-quality one-on-one instruction can significantly improve educational outcomes [2], [3], [4]. However, in many Low and Middle-Income Countries (LMICs), including West Africa, the low supply and high cost of quality tutors mean that many learners cannot access this type of support [5]. Thus, research is needed to identify affordable and feasible interventions appropriate for this space.

In recent years, interest has been increasing in technology-enabled approaches such as adaptive learning environments and virtual tutors to offer additional support to learners where personal tutors are not available. These approaches have been used to varying degrees of success in supplementing traditional instruction, but, when they work, they are extremely cost-effective, a consideration that is particularly important in LMIC contexts. However, a major hurdle to extending the use of these types of tools to LMICs, has been the lack of widespread access to personal computers and reliable, high-speed internet that these learning environments typically require. In West Africa, for instance, less than 20% of the population has access to home computers and robust internet connections [6]. However, mobile phone usage is remarkably high in the region, with data indicating that around 90% of West Africans own a mobile phone and reside in areas with 3G coverage [6], [7]. This trend suggests the possibility of implementing interactive educational experiences on mobile platforms, circumventing the need for expensive personal computers and high-speed internet.

Rising Academies, an educational network based in Ghana, has created Rori, a free AI-powered math tutor available on WhatsApp. Inspired by successful elements of Teaching at the Right Level and Intelligent Tutoring System, Rori has students work their way through a curriculum of over 500 micro-lessons, based on the Global Proficiency Framework, an internationally recognized set of math learning standards. Each micro-lesson includes a brief explanation and a series of scaffolded practice questions tied to a specific learning standard. If a student struggles with a question, they receive a hint first and then a worked solution.

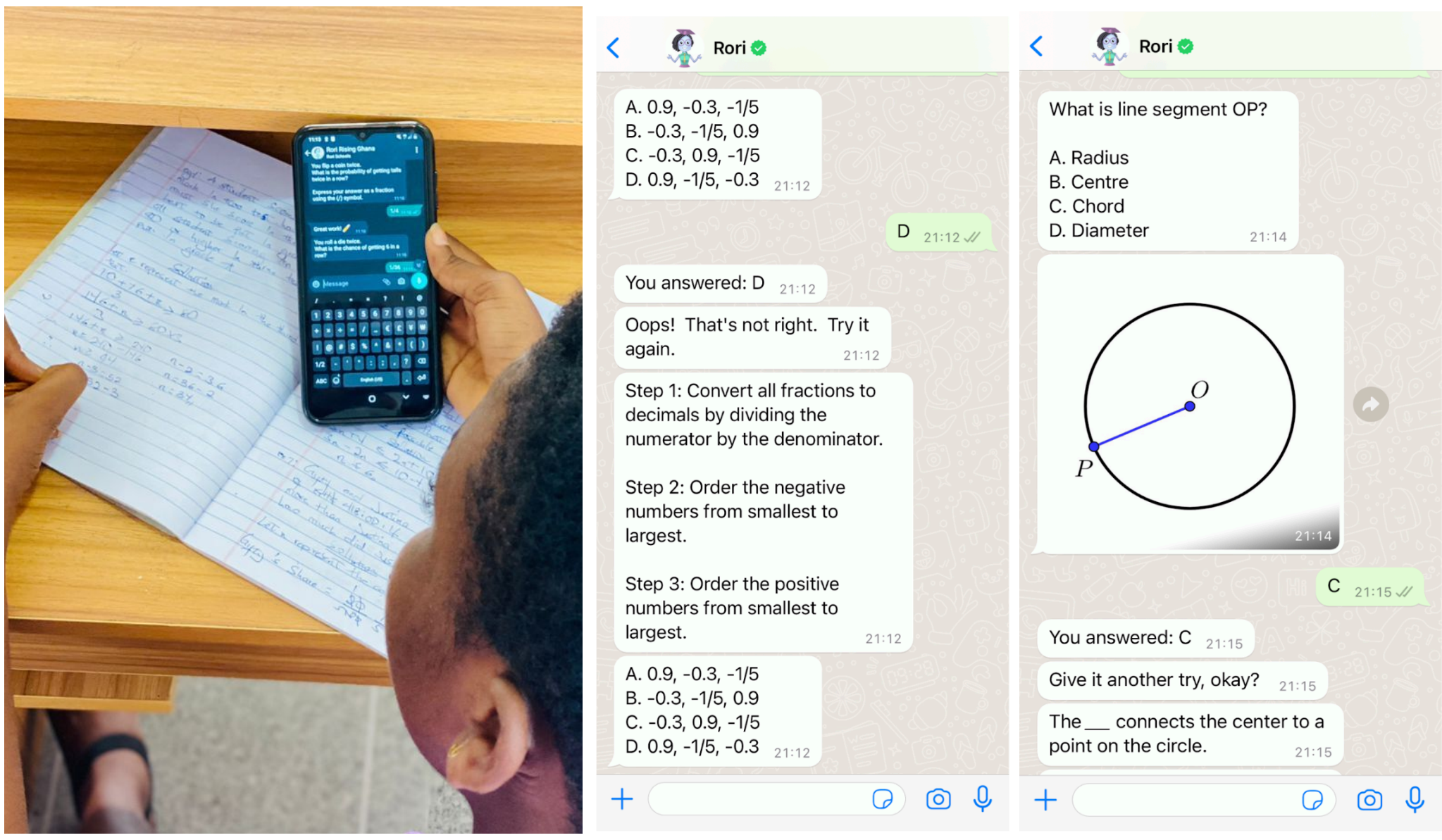

**Fig. 1.** Interaction with Rori while  solving math problems.

A key component of Rori is the use of various natural language processing (NLP) methods, including specialized language models (LLMs). This enables students to chat with Rori using natural language, rather than simple navigating a pre-defined button-based dialog. The natural language interface of Rori positions it to both better interpret students' answer attempts and questions, and to engage students in more open-ended conversations about math goals and metacognitive skills. For more context you can **watch this 2-minute video**.

To evaluate if receiving extra math instruction provided by Rori impacts math learning, students in grades 3-9 from Rising Academies' schools in Ghana participated in this study. The schools were divided into treatment and control groups. The treatment group, as a supplement to regular math instruction, engaged in two 30-minute weekly sessions with Rori during their study hall period. The control group continued regular math instruction and did not receive access to Rori during study hall. Students in both groups were given a math assessment twice over 8 months which made it possible to assess the impact of Rori on math performance while controlling for confounding variables through the comparison of similar schools (i.e., a difference-in-difference measure between the two groups).

## 2 Prior Work

### 2.1 ITSs and Adaptive Learning Environments

While high-dosage tutoring is a highly effective tool for enhancing student learning [2], [4], its scalability is hindered by operational and financial constraints. Advancements in technology have led to attempts to replicate this effect by delivering personalized education through Intelligent Tutoring Systems (ITSs) which aim to adapt the learning experience to individual student needs, customizing the content and pace, thus emulating one-on-one tutoring [8]. These systems have been implemented in various educational contexts, such as K-12 classrooms and universities, and there is evidence that they can be effective, particularly for the teaching of mathematics and science [9]. In addition to ITSs, other related approaches have emerged in the field of educational technology. Adaptive learning systems, for example, use algorithms to tailor content and pace to students' needs and typically allow students to move through the curriculum in a flexible and personalized manner [10]. While opinions vary on their effectiveness, these approaches hold potential in addressing several educational challenges, particularly in the context of resource-constrained education systems where increasing instructional time, or providing high-dosage tutoring would not be possible [11].

### 2.2 Teaching at the Right Level

Despite a successful push to increase school enrolment in LMICs, this has not always translated into improvements in learning for all students[1]. Practically, many national curricula primarily cater to high-achieving students, neglecting most children who lag behind. Various factors at both school and home contribute to this predicament [12]. One effective approach to this challenge has been Teaching at the Right Level (TaRL) [13]. First, children's learning levels are evaluated, then they are grouped based on those levels rather than age or grade, and taught at their current ability level, using a curriculum that prioritizes foundational skills [12]. Randomized evaluations conducted in India over 2001-2020, Western Kenya in 2005, and Botswana in 2020 consistently show that programs employing these strategies improve learning outcomes. For example, research by Nobel laureates Abhijit Banerjee and Esther Duflo demonstrated a 0.28

SD improvement in learning outcomes, equivalent to an additional 1.5 years of schooling [13]. While extremely effective, inventions such as TaRL typically require a non-trivial amount of teacher training as well as operational changes in a school setting, which are challenges to wide-spread adoption. As result, there has been interest in seeing if technology-supported inventions can partially emulate the successful components.

### 2.3 Technology Supported Learning in LMICs

The personalization inherent in Intelligent Tutoring Systems has substantial similarities with the TaRL approach, and some interventions have tried to leverage technology to implement individualized teaching strategies [13], [14]. A study of a computer-assisted intervention using similar principles resulted in a 0.47 SD improvement, approximately 2.5 years of schooling, and a recent randomized control trial on a computer-assisted program estimated a 0.6 SD improvement in math scores over a year, or approximately an additional three years of schooling [14]. However, students in resource-constrained settings often don't have access to home computers and the internet, prompting exploration for more cost-effective and accessible educational solutions.

There have been attempts to explore whether phone-based interventions can offer a viable alternative for delivering educational content and support, particularly for cases of disruption in traditional educational settings [15]. For instance, in Kenya, M-Shule offers SMS-based micro-courses and personalized tuition tools to support primary school students in English, Kiswahili, and Math. [16]. However, other studies found that another mobile-based educational intervention did not produce learning gains [17]. Considering the diverse findings, our research seeks to contribute to the growing body of literature on mobile learning in LMICs by examining the impact of Rori on math learning gains. We aim to shed light on the potential of such interventions to provide a cost-effective and accessible means of delivering educational content and support to students in these low-resource settings.

## 3 Current Study

### 3.1 Overview

Students from 11 of Rising Academies' schools in Ghana were involved in the study. These schools were selected based on their similarities in geography, demographics, curricula, and teaching methodologies. Schools were randomly assigned to two groups: a control group, which included students from six schools; and a treatment group, which included students from five schools. The assignment was done as the school level, as assignment at the student level was not feasible for a variety of reasons.

The treatment group participated in two 30-minute sessions with Rori every week, during their study hall period. While teachers were present during these sessions to help resolve potential technical issues, and students spent the period working independently on Rori; **watch a clip of a Rori session**. Meanwhile, the students in the control group did not have access to Rori during their study hall period and continued to receive their regular math instruction. Both groups completed math assessment in early February (baseline) and late August (endline) of 2023. The same math assessment was used at both timepoints and all grade levels. The assessment consisted of 35 questions, each worth one point, with a mixture of multiple-choice and open-response questions covering numeracy and algebra skills from grades 3 to 5 on the Global Proficiency Framework.

Rori is an intervention that is characterized by its low cost and minimal disruption to school operations. Other potentially efficacious interventions such as extended instructional time, smaller classroom sizes, or high-dosage tutoring would simply not be feasible due to the associated costs. Resultingly, we compare the incremental impact of Rori relative to normal schooling, which allows us to understand the specific impact of potentially scalable intervention in enhancing educational outcomes.

### 3.2 Participation

Initially, 637 students in grades 3-8 participated in the baseline assessment and out of that group, 477 students also completed the endline. The attrition of 160 students was primarily due to inconsistent school attendance; it is common for students to miss school periodically due to personal reasons. Of the 477 students who participated in both tests, 241 were in the control group and 236 were in the treatment group. Statistical examinations were conducted to assess any systematic differences between the control and treatment group at baseline and to determine if there were distinctions between students who completed the study and those who did not. We found no systematic difference in control and treatment group, and only small differences in the students who took the baseline but not the endline, which is discussed in detail in the subsequent section.

## 4 Results

### 4.1 Estimated Learning Gains

To assess the impact of using Rori on learning, growth scores were computed by subtracting baseline raw scores, the number of questions answered correctly, from endline raw scores for each student who completed both tests. An independent samples t-test between the control (M = 2.12, SD = 6.30) and the treatment group (M = 5.13, SD = 7.03) revealed that the 3.01 difference in growth scores was highly statistically significant ($p < 0.001$).

**Table 1.**
Descriptive statistics for each condition.

| Condition | Control (N=241) | | Treatment (N=237) | |
|---|---|---|---|---|
| Test | Baseline | Endline | Baseline | Endline |
| Mean | 20.20 | 22.32 | 20.29 | 25.42 |
| St Dev | 8.81 | 8.06 | 8.72 | 7.25 |
| Growth | 2.12 | | 5.13 | |

To understand the magnitude of the difference in learning gains between the treatment and control groups, we calculated the effect size using the pooled standard deviation from both conditions and time points, as suggested by Morris [18]. The calculated effect size between the baseline and endline, expressed as Cohen's d, was found to be 0.36. This effect size would be considered a moderate to large effect size in educational research. Hattie et al. would categorize 0.29 as moderate, and similar to the magnitude of the effect of a good teacher [19]. Whereas Kraft proposes argued that educational interventions with an 0.20 SD of over should be considered as large [20].

Another way to get a sense of the magnitude of the learning gains is to compare the learning gains of the control group (2.12) – which is indicative of the amount of improvement due to normal classes and potential retest effects – to that of the treatment group (5.13), which suggests the differential impact of Rori might be equal or greater than a year of schooling.

### 4.2 Comparing Characteristics of Treatment and Control Group

To establish baseline equivalence and ensure comparability of the two groups, baseline scores, gender and age were compared. No statistically significant difference was found in the baseline scores of the two groups ($t(475) = -0.17$, $p = 0.87$), suggesting that the two groups had equivalent levels of math knowledge at the start of the evaluation.

Gender distribution was compared across the two groups. The chi-square test for the association between gender and condition yielded a non-significant result ($\chi^2 = 0.93$, $p = 0.33$), suggesting no statistically significant difference in gender distribution between the control and treatment groups. Age variation was also examined, and an independent samples t-test was conducted, and the results indicated no statistically significant difference in mean age between the control and treatment groups ($t\ (475) = 1.23$, $p = 0.22$).

Overall, these analyses indicate that the control and treatment groups had statistically similar baseline scores and were balanced across age and gender. The similarity of the control and treatment group combined with the fact that the results report a difference in difference score (i.e., a “growth” score), suggest that the improvement in learning observed is attributable to the intervention (i.e., using Rori) rather than other measured factors (e.g., prior aptitude, gender, age).

### 4.3 Comparing Students Who Completed Both Tests vs Those Who Did Not

The baseline was taken by 637 students in grades 3-8 (336 in Treatment, 301 in Control), of which 477 took the endline and 160 did not. The mean age for the students who took both tests was 12.10 (SD = 1.95) and those who dropped out was 11.93 (SD = 2.15) and there was no statistically significant difference between the two groups ($t(635) = 0.88$, $p = 0.38$). The gender distribution was compared between the students who completed both assessments and those who dropped out. The chi-square test for the association between gender and test completion yielded a non-significant result ($\chi^2 = 0.766$, $p = 0.38$).

The mean raw baseline score of those that completed both tests was 22.26 (SD = 7.57) and the mean score of those that dropped out was 18.53 (SD = 7.70). While there was a difference in baseline ability between these two groups of students, the reported effect size of the intervention is calculated based on growth scores of students who completed both tests, so it would not be impacted by this difference. None-the-less, this difference needs to be investigated as there are many possible explanations that are not within the scope of this paper.

## 5 Discussion

### 5.1 Implications

The initial evidence reveals that receiving extra math practice with Rori led to substantial learning gains across multiple grade levels. The intervention's effectiveness in improving math skills is not confined to a narrow age band but is instead spread across a broad spectrum, indicating that it can adapt to different learning stages without the need for considerable financial investment. While the results should be interpreted with caution, they do suggest that chat-based tutoring solutions leveraging AI could offer a cost-effective and operationally efficient approach to enhancing learning outcomes for students globally.

In a policy context, the effect size of an intervention like Rori is favorable in relation to its cost to implement. The cost structure (i.e. LLM API, mobile device, 3G data costs) presents a scalable model, with initial setup expenses offset by the possibility of achieving an annual cost as low as $5 per student. Therefore, it could be a financially viable solution for resource-constrained educational settings. In addition to cost, interventions can struggle to scale because of the difficulties of replicating technical and support infrastructure in new geographies. Relative to changing curricula or teaching practices, Rori is relatively straightforward to implement once schools have access to mobile phones, as it does not require additional teaching staff or extensive training. Relying on widely used and accessible devices, such as budget phones and chat platforms like WhatsApp, minimizes dependence on existing infrastructure, avoiding the limitations posed by the lack of personal computers and reliable high-speed internet.

### 5.2 Limitations

While the experimental design of this study was robust and the learning gains were both large and statistically significant, there are certain limitations. With educational interventions there is always a potential for unobserved differences between the treatment and control group. While schools were randomly assigned to treatment and control groups and both had the same curriculum, lesson plans, and timetable, it is possible that the school administrators and teachers at the treatment schools changed their behavior because of increased observation (i.e. Hawthorne effect). However, schools in the Rising Academies Network are accustomed to routine testing and observation, so this is less likely to be a potential confound. Relatedly, while multiple analyses were conducted to investigate baseline equivalence of the two groups, there may have been unobserved differences in students' prior ability, or subtle differences in their socio-economic background that could have influenced the results.

Another potential limitation was the design of the assessment. To make tracking student progress comparable across grades, the same assessment was used for all students. This led to some students, particularly those in higher grade levels, receiving perfect scores at baseline, so it was not possible to observe their improvement over time. While this would suggest that growth scores might have been *understated*, it is also possible that Rori is only effective in helping students with relatively easier math topics, and that the learning gains might be diminished if older students were tested on more difficult topics.

### 5.3 Further Research

There are several opportunities for future research, as mentioned an examination of the observed attrition in this study is warranted. Another area of investigation that was outside the scope of this study was the specific relationship between time spent using Rori and improved math ability. It is plausible that student learning was an approximately linear function of the amount of time students spent using Rori, but it is equally plausible that the relationship is nonlinear and that the beneficial effect of Rori plateaus after a certain amount of usage. Subsequent research investigating dosage over time might also reveal interactions with learning gains, such as identifying an optimal amount of interaction time before reaching diminishing returns in performance or potential interference with learning in other subjects. Future studies could also employ more diverse and comprehensive assessment tools, reducing the ceiling effects observed in this study and enhancing the validity and reliability of student evaluation. Finally, future studies could investigate the impact of Rori in other schools outside of Rising Academies Network or in other countries. Rori could also be used by children at home. This would contribute to the generalizability of this study's findings that using Rori improves math learning.

## 6 Conclusion

The present study explored the impact of the artificial intelligence conversational math tutor accessible through WhatsApp on the math performance students in Ghana. We found that the treatment group's math growth scores were markedly higher, with an effect size of 0.36, and data exhibiting statistical significance ($p < 0.001$). Given its ability to operate on basic mobile devices on low-bandwidth data networks, the intervention exhibits substantial potential in supporting personalized learning in other LMICs. In such regions, the possession of laptops and access to high-speed internet connectivity, requisites for many video-centered learning platforms, remain significantly limited.

However, it is prudent to interpret these results with caution. As this study provides insight from just the first year of intervention in a school-based context, it underlines the necessity for additional, future research to ascertain the conditions essential for ensuring its successful implementation, but also identifies the importance of a personalized, adaptive approach to enrich students' learning experience. Nevertheless, the results presented here highlight the potential of chat-based tutoring solutions leveraging artificial intelligence to offer a cost-effective strategy to boost learning outcomes for disadvantaged students globally.

## References

[1] World Bank, UNESCO, UNICEF, USAID, FCDO, Bill & Melinda Gates Foundation, "The State of Global Learning Poverty: 2022 Update," 2022.

[2] B. S. Bloom, "The 2 Sigma Problem: The Search for Methods of Group Instruction as Effective as One-to-One Tutoring," *Educ. Res.*, vol. 13, no. 6, p. 4, Jun. 1984, doi: 10.2307/1175554.

[3] P. F. Vadasy, J. R. Jenkins, L. R. Antil, S. K. Wayne, and R. E. O'Connor, "The Effectiveness of One-to-One Tutoring by Community Tutors for at-Risk Beginning Readers," *Learn. Disabil. Q.*, vol. 20, no. 2, pp. 126–139, May 1997, doi: 10.2307/1511219.

[4] M. T. H. Chi, S. A. Siler, H. Jeong, T. Yamauchi, and R. G. Hausmann, "Learning from human tutoring," *Cogn. Sci.*, vol. 25, no. 4, pp. 471–533, Jul. 2001.

[5] Bray, M., "Shadow education in Sub-Saharan Africa: scale, nature and policy implications." 2021. [Online]. Available: https://unesdoc.unesco.org/ark:/48223/pf0000380073

[6] "Overview of state of digital development around the world based on ITU data." International Telecommunication Union (ITU), 2022. [Online]. Available: https://www.itu.int/en/ITU-D/Statistics/Dashboards/Pages/Digital-Development.aspx

[7] L. Silver and C. Johnson, "Internet Connectivity Seen as Having Positive Impact on Life in Sub-Saharan Africa." Pew Research Center.

[8] K. Vanlehn, "The Relative Effectiveness of Human Tutoring, Intelligent Tutoring Systems, and Other Tutoring Systems," *Educ. Psychol.*, vol. 46, no. 4, Art. no. 4, Oct. 2011.

[9] S. Steenbergen-Hu and H. Cooper, "A meta-analysis of the effectiveness of intelligent tutoring systems on college students' academic learning.," *J. Educ. Psychol.*, vol. 106, no. 2, pp. 331–347, May 2014, doi: 10.1037/a0034752.

[10] P. Phobun and J. Vicheanpanya, "Adaptive intelligent tutoring systems for e-learning systems," *Procedia - Soc. Behav. Sci.*, vol. 2, no. 2, pp. 4064–4069, 2010. d

[11] B. D. Nye, "Intelligent Tutoring Systems by and for the Developing World: A Review of Trends and Approaches for Educational Technology in a Global Context," *Int. J. Artif. Intell. Educ.*, vol. 25, no. 2, pp. 177–203, Jun. 2015, doi: 10.1007/s40593-014-0028-6.

[12] Abdul Latif Jameel Poverty Action Lab (J-PAL), "'Teaching at the Right Level to improve learning.' J-PAL Evidence to Policy Case Study." 2018.

[13] Banerjee, Abhijit V. and Banerji, Rukmini and Berry, James and Duflo, Esther and Kannan, Harini and Mukerji, Shobhini and Shotland, Marc and Walton, Michael, "Mainstreaming an Effective Intervention: Evidence from Randomized Evaluations of 'Teaching at the Right Level' in India," *MIT Dep. Econ. Work. Pap. No 16-08*, Nov. 2016.

[14] K. Muralidharan, A. Singh, and A. Ganimian, "Disrupting Education? Experimental Evidence on Technology-Aided Instruction in India," *Am. Econ. Rev.*, vol. 109, no. 4, pp. 1426–60, 2019.

[15] R. F. Kizilcec, M. Chen, K. K. Jasińska, M. Madaio, and A. Ogan, "Mobile Learning During School Disruptions in Sub-Saharan Africa," *AERA Open*, vol. 7, p. 233285842110148, Jan. 2021, doi: 10.1177/23328584211014860.

[16] Lancaster University, Lancaster, UK and K. Jordan, "Learners and caregivers barriers and attitudes to SMS-based mobile learning in Kenya," *Afr. Educ. Res. J.*, vol. 11, no. 4, pp. 665–679, Dec. 2023, doi: 10.30918/AERJ.114.23.088.

[17] R. F. Kizilcec and M. Chen, "Student Engagement in Mobile Learning via Text Message," in *Proceedings of the Seventh ACM Conference on Learning @ Scale*, Virtual Event USA: ACM, Aug. 2020, pp. 157–166. doi: 10.1145/3386527.3405921.

[18] S. B. Morris, "Estimating Effect Sizes From Pretest-Posttest-Control Group Designs," *Organ. Res. Methods*, vol. 11, no. 2, pp. 364–386, Apr. 2008, doi: 10.1177/1094428106291059.

[19] J. Hattie, *Visible learning: a synthesis of over 800 meta-analyses relating to achievement*, Reprinted. London: Routledge, 2010.

[20] M. A. Kraft, "Interpreting Effect Sizes of Education Interventions," *Educ. Res.*, vol. 49, no. 4, pp. 241–253, May 2020, doi: 10.3102/0013189X20912798.